\documentclass[11pt,twoside]{article}

\usepackage{asp2006}
\usepackage{epsf}
\usepackage{psfig}
\usepackage{lscape}
\usepackage{graphicx}

\begin{document}
\title{Exoplanetary Systems with SAFARI:\\ A Far Infrared Imaging Spectrometer for SPICA} 

\author{J.~R. Goicoechea$^1$ and B. Swinyard$^2$\\ on behalf of the SPICA/SAFARI science teams}                                   
\affil{$^1$Centro de Astrobiolog\'{\i}a (CSIC-INTA), Madrid, Spain} 

\affil{$^2$Rutherford Appleton Laboratory, Chilton, Didcot, UK}    


\begin{abstract} 
The far-infrared (far-IR) spectral window plays host to a wide range of  spectroscopic 
 diagnostics with which to study planetary disk systems and exoplanets 
at wavelengths completely blocked by the Earth  atmosphere.
These include the thermal emission of dusty belts in debris disks, the water ice 
features in the ``snow lines" of protoplanetary disks, as well as
many key chemical species (O, OH, H$_2$O, NH$_3$, HD, etc). 
These tracers play a critical diagnostic role in  a number  of key areas including 
the early stages of planet formation and potentially, exoplanets. 
The proposed Japanese-led IR space telescope SPICA, 
with its 3m--class cooled  mirror ($\sim$5\,K) will be the next step in sensitivity after 
\textit{ESA's Herschel Space Observatory} (successfully launched in~May 2009). 
SPICA is a candidate ``M-mission" in  \textit{ESA's Cosmic Vision~2015-2025} process.
We summarize the science possibilities of SAFARI: 
a far-IR imaging-spectrometer (covering the $\sim$34 -- 210\,$\mu$m band) that is one of 
a suite of instruments for SPICA.
\end{abstract}

\section{A New Window in Exoplanet Research}   
The study of exo-planets (EPs) requires many different approaches across the full wavelength 
spectrum to both discover and characterize the newly discovered objects in order that
 we might fully understand the prevalence, formation, and evolution of planetary systems.  
The mid--IR and far--IR spectral regions
are especially important in the study of planetary atmospheres as it spans both the peak of 
thermal emission from the  majority of EPs thus far discovered (up to $\sim$1000 K) 
and is particularly rich in molecular  features that can uniquely identify the chemical 
composition,  from protoplanetary disks to planetary atmospheres and trace the
 fingerprints of primitive biological activity.  In the coming decades many space- and 
ground-based facilities are planned that are designed to search for EPs
on all scales from massive, young ``hot Jupiters", through large rocky super-Earths 
down to the detection of exo-Earths. 
Few of the planned facilities, however, will have the ability to 
characterize the planetary atmospheres which they discover through the application of mid-IR
and far-IR spectroscopy.  SPICA  will be realized within 
$\sim$10 years and 
has a suite of instruments that can be applied to the detection and characterization of 
EPs over the $\sim$5--210\,$\mu$m spectral range 
(see e.g.\ our \textit{White Paper}; Goicoechea et al. 2008). 

\begin{figure}[ht]
  \centering 
  \includegraphics[height=0.99\hsize{},angle=-90]{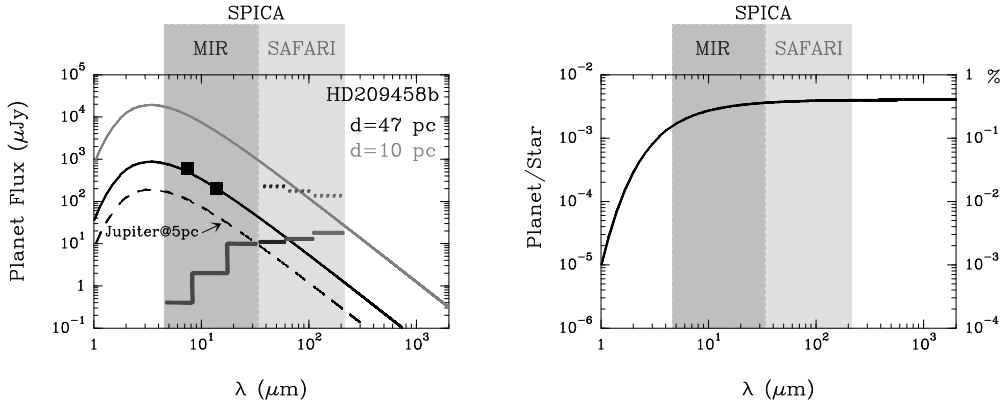} %
   \caption{\textbf{Left}: Fit to HD 209458b ``hot Jupiter" (T$_{\rm eff} \simeq$\,1000~K) fluxes inferred  
from a secondary transit  with \textit{Spitzer} \citep{swai08}
around a G0 star ($d \sim$\,47~pc, in black) and interpolation to  
$d$\,=\,10~pc (gray). The emission of a cooler Jupiter-like planet at 5\,pc is 
shown in dashed (reflected emission neglected). 
Thick horizontal lines are the 5$\sigma$-1hr photometric sensitivities of 
SPICA mid-IR instruments  and SAFARI.
Dotted lines show sensitivities in spectrophotometric mode  ($R\simeq25$).  
SPICA will observe similar transits of \textit{inner} ``hot Jupiter" routinely and will potentially
 extract their IR spectrum  (rich in H$_2$O, O$_3$, CH$_4$, NH$_3$ and HD features as in Solar System planets). 
\textbf{Right}: Increasing planet-to-star contrast at longer mid-- and far--infrared wavelengths \citep{goi08}.}
  \label{fig:safari-sensitivity}
\end{figure}

The SAFARI instrument \citep{swi09} will provide capabilities to complement SPICA studies in the mid-IR
(either coronagraphic or transit studies). Indeed, SAFARI could be the only planned instrument 
able to study EPs in a completely new wavelength domain (for SAFARI's band 1)
not covered by \textit{JWST} nor by \textit{Herschel} (Fig.~1). 
This situation is often associated with
unexpected discoveries.    Since cool EPs   show much higher contrast in the far-IR than in the near--/mid--IR 
(e.g.\ Jupiter's effective temperature is $\sim$\,110~K), if 
such EPs are found in the next 10 years, their transit studies with SPICA will help to 
constrain their main properties, which are much more difficult to infer at shorter wavelengths. 
Note that following \textit{Infrared Space Observatory} (ISO) observations, Jupiter seen at 5~pc will produce
 a flux of 
$\sim$\,35~$\mu$Jy at 37~$\mu$m, but less than $\sim$\,1~$\mu$Jy  at 15~$\mu$m.
 SAFARI band1 ($\sim$\,34~--60~$\mu$m) hosts a variety of interesting atmospheric molecular 
features (e.g.\ H$_2$O at 39~$\mu$m, HD at 37~$\mu$m and NH$_3$ at 40 and 42~$\mu$m). 
Strong emission/absorption of these features was first detected by ISO in the atmospheres 
of Jupiter, Saturn, Titan,  Uranus and Neptune \citep{feuc99}. 

\acknowledgements 
We warmly thank the SPICA/SAFARI science
teams for fruitful discussions. JRG is supported by a \textit{Ram\'on y Cajal} research contract.

\end{document}